\documentstyle[preprint,aps,epsf]{revtex}
\tighten
\draft
\widetext
\preprint{UPR-942-T, NSF-ITP-01-65, CLNS-01/1743}
\bigskip
\bigskip
\begin{document}
\title{Some Aspects of Brane Inflation}
\medskip
\author{Gary Shiu$^{1,2}$\footnote{Electronic mail:
shiu@dept.physics.upenn.edu}
and S.-H.~Henry Tye$^3$\footnote{Electronic mail: tye@mail.lns.cornell.edu}}
\address{$^1$ Department of Physics and
Astronomy, University of Pennsylvania, Philadelphia, PA 19104 \\
$^2$ Institute for Theoretical Physics, University of California,
Santa Barbara, CA 93106 \\
$^3$ Newman Laboratory of Nuclear Studies, Cornell University, Ithaca,
NY 14853}
\date{June 28, 2001}
\bigskip
\medskip
\maketitle

\def\kl#1{\left(#1\right)}
\def\th#1#2{\vartheta\bigl[{\textstyle{  #1 \atop #2}} \bigr] }

\begin{abstract}

{} The inflaton potential in four-dimensional theory is rather arbitrary, 
and fine-tuning is required generically. By contrast, inflation in the 
brane world scenario
has the interesting feature that the inflaton potential is motivated from
higher dimensional gravity, or more generally, from bulk modes or string 
theory.
We emphasize this feature with examples. We also consider 
the impact on the spectrum of density perturbation
from a velocity-dependent potential between branes 
in the brane inflationary scenario. It is likely that such a potential
can have an
observable effect on the ratio of tensor to scalar perturbations.

\end{abstract}

\section{Introduction}

By now, the inflationary universe \cite{guth,newinflation} 
is generally recognized to
be a likely scenario that leads to the big bang. So far, its predictions of 
the flatness and the scale-invariant power spectrum of the density 
perturbation that seeds structure formations are in good agreement 
with observations \cite{new}. Future data will test inflationary
predictions to a very high accuracy. In fact,
there is widespread hope that
the enormous growth in scale during inflation may provide a 
kind of Planck scale microscope, allowing us one day to probe
stringy and/or brane world effects from
precision cosmological measurements \cite{EGKS}.

In conventional inflationary models, the physics lies in the inflaton 
potential. Although there are constraints from observations,
such as enough number of e-foldings, the amplitude of the density 
perturbation etc., there are very few theoretical constraints on the 
underlying dynamics of the inflaton field.
As a result, this scalar 
field potential can take a large variety of shapes. In hybrid inflation 
\cite{Linde} or other variations, additional fields and/or parameters 
are introduced, allowing even more freedom.  
However, the problem in a conventional 4-dimensional theory is the 
difficulty of finding an appropriate inflaton potential. In general, 
a potential that yields the correct magnitude of density perturbation
and satisfies the slow-roll condition is not well-motivated and requires 
some fine-tuning.
Generically, such fine-tuning is not preserved by quantum corrections. 

Recently, motivated by the idea of brane world \cite{add,ST,BW,ovrut}, 
the brane inflationary scenario\cite{dvali-tye} was proposed, where the 
inflaton is identified with an inter-brane separation: 
${\bf \Phi} = M_s^2 {\bf r}$. Inflation ends when the branes collide, heating 
the universe that starts the big bang.
In this scenario, the inflaton potential has a geometric interpretation. 
In particular, higher dimensional gravity (or more generally, bulk modes 
or closed string states) dictates the form of the inflaton potential.
For example, the inflaton potential may simply be the Newton's potential 
between branes. 
This visualization of the brane dynamics allows one to implement inflation 
physics pictorially. Suitable inflaton potentials naturally appear in the 
brane world scenario, and their forms are robust under quantum corrections.
In this paper, we demonstrate further this unusual feature of brane inflation.
If the fundamental string scale is substantially above a few TeV, brane
inflation will be a valuable testing ground for the brane world scenario.

The motion of the branes is dictated by interbrane 
(which can be brane-brane, 
brane-orientifold or brane-antibrane) forces. 
Moreover, a velocity-dependent term in the potential is present generically.
So when branes move, this term may become important.
This velocity-dependent term is calculable in string theory;
the precise values of the parameters that appear in this term
depend on the details of the model.
In fact, one might expect such a velocity-dependent term 
from the post-Newtonian approximation in higher dimensional 
gravity.
In this paper, we also examine the impact of such a velocity-dependent 
force on the slow-roll and the power spectrum of the density perturbation.
We find that the effect on the tilt of the
spectral index of the density perturbations
may be small. However, such a term is likely to have an observable effect
on the ratio of the tensor to scalar perturbations.
Moreover, 
the scale dependence of the spectral index is also modified. Therefore,
a global analysis of data from various measurements
may allow us to distinguish these effects from that of 
the conventional inflationary scenario.

\section{Dynamics of the Inflaton}
 
Our starting part is the effective action $\Gamma (\phi)$
for the canonically normalized
inflaton $\phi$ of the form:
\begin{equation}\label{effective}
\Gamma (\phi) = \int d^4 x \sqrt{-g} \left( -V (\phi)
+ \frac{1}{2} Z (\phi) \partial_{\mu} \phi \partial^{\mu} \phi
+ \dots  \right)
\end{equation}
where $V$ is the effective potential. 
The precise form
of $V (\phi)$ will be specified later on. 
Note that the field $\phi$ in the brane world is related to the 
separation $r$ between the 
branes by $\phi=M_s^2 r$, where $M_s$ is the string scale. 
To lowest order $Z (\phi)=1$. There are also higher order corrections
to $Z (\phi)$. In the usual inflationary scenario, the deviation of $Z(\phi)$
from unity is due to quantum effects. Typically, it takes the 
form of $Z(\phi) \sim 1 + c ~g^2 ~\log (\phi)$, where
$c$ is a constant and $g$ is the coupling constant.
Here, we will consider 
a different type of contribution to $Z(\phi)$ which is motivated by brane
world physics (or higher dimensional gravity).

In an expanding universe, $ds^2=dt^2-a^2(t) d \vec{x}^2$, then,
\begin{equation}
\Gamma (\phi) = 
\int d^4 x \left( -a^3 V(\phi)+\frac{1}{2} {a}^3 Z(\phi) \dot{\phi}^2 
- \frac{1}{2} a Z(\phi) (\nabla \phi)^2 \right)
\end{equation}

For inflation to take place, there is some choice of $\phi$ so that the
potential $V(\phi)$ satisfies the slow-roll conditions:
\begin{eqnarray}
\label{slow-roll}
\epsilon &=& \frac{1}{2} M_P^2 \left( \frac{V^{\prime}}{V} \right)^2  << 1
 \nonumber \\ 
|\eta|  &=&  |M_P^2 \frac{V^{\prime\prime}}{V}| << 1
\end{eqnarray}
where prime indicates derivative with respect to $\phi$.
The spatially inhomogeneous terms will be redshifted away by inflation,
so they are ignored at this stage. 
For slow-roll, we also expect
\begin{equation}\label{inflation-cond}
\frac{1}{2} Z (\phi) \dot{\phi}^2 << V(\phi) ~.
\end{equation}
as long as $Z(\phi)$ is not too far from unity.
The slow-roll conditions imply that (\ref{inflation-cond}) is satisfied.
The equation of motion for $\phi$ is
\begin{equation}\label{eom}
Z \left( \ddot{\phi} + 3 H \dot{\phi} 
 \right) - \frac{1}{2} Z^{\prime} \dot{\phi}^2 
+ V^{\prime} = 0
\end{equation}
where the Hubble constant is given by, during the slow-roll epoch,
\begin{equation}
H^2 = \left(\frac{\dot{a}}{a} \right)^2 = \frac{V}{3 M_P^2}
\end{equation}
With the condition $\frac{1}{2} Z^{\prime} \dot{\phi}^2 << V^{\prime}$,
we have:
\begin{equation}\label{dotrp=2}
\dot{\phi} \sim - \frac{V^{\prime}}{3 H Z(\phi)}
\end{equation}

The fluctuation $\delta \phi (\vec{x})$ results in a position-dependent
time delay in the evolution of $\phi$, which in turn gives rise to
the density fluctuation. A crucial step in this time-delay
formalism \cite{GuthPi} is the observation that in the de Sitter
case,
$\delta \phi$ and $\dot{\phi}$ obeys the same time dependent
differential equation when the scale is outside the horizon. This allows
us to write 
$\delta \phi (\vec{x},t) \sim - \delta \tau (\vec{x}) \dot{\phi} (t)$,
where $\tau(\vec{x})$ is time-independent. 
The fourier modes $|\delta_k|$ of 
the density perturbation $\delta \rho/\rho$ 
can then be expressed in terms
of the quantum fluctuations $\delta \phi_k$ and $\dot{\phi}$
(see Appendix):
\begin{equation}\label{gpformula}
|\delta_k| = H \frac{\delta \phi_k}{\dot{\phi}}
\end{equation}
where $k$ is the comoving wavenumber.
In a more general setting, we can apply 
the gauge-invariant formulation \cite{DensityPerturbation} to 
calculate unambiguously the density perturbation. 

We note 
that after the scale crosses the horizon (hence the $\nabla \phi$ term
for $\delta \phi$ is unimportant),
$\delta \phi$ obeys the same
equation as $\dot{\phi}$
in brane inflation, even the velocity-dependent potential
is included.
In the de Sitter case, $\delta \phi_k$ is given by
$\delta \phi_k = {H}/{2 \pi}$.
Then
\begin{equation}\label{delta_k-vel}
|\delta_k| = \frac{3H^3 Z}{2 \pi V'}
\end{equation}
since the velocity of the inflaton is modified.
We can consider a more general case where the factor 
$Z$ is raised to a power.

In the absence of velocity-dependent potential, the ratio of tensor to 
scalar fluctuations is predicted by the inflationary 
scenario \cite{Starobinsky}.
Here, we note that the velocity dependent potential affects only the 
scalar fluctuation but not the tensor perturbation. The reason
is that the scalar and tensor fluctuation have different origins.
The scalar fluctuation comes from the quantum fluctuation of a brane
mode, and hence from Eq.(\ref{dotrp=2}) and the fact that $\delta \phi_k$ is
not modified (see Appendix), the amplitude $\delta_S^2$ is increased by 
a factor of $Z^2$.
(We denote $\delta_k$ for the
scalar perturbation by $\delta_S$).
On the other hand, the tensor fluctuation arises from the fluctuation of 
the bulk (or closed string) modes which is independent of $Z$, 
hence the fourier modes of the tensor perturbations $\delta_T$:
\begin{equation}
\delta_T^2 \sim \frac{H^2}{M_P^2} \sim \frac{V}{M_P^4}
\end{equation}
is not affected by (\ref{dotrp=2}).
Therefore, we expect the prediction of \cite{Starobinsky}
to be modified:
\begin{equation}\label{T/S}
\frac{\delta_T^2}{\delta_S^2} = 12.4 ~\epsilon
\quad \longrightarrow \quad 
\frac{\delta_T^2}{\delta_S^2} = 12.4 ~\frac{\epsilon}{Z^2}
\end{equation}
As a result, the ratio of tensor to scalar perturbations by itself
does not give a measurement of $\epsilon$. 
As an illustration, let us take $Z \sim 0.7$, 
the ratio of tensor to scalar perturbations is increased by a factor 
of $2$.

Let us now consider the effect on the density perturbation power 
spectrum index $n$ due to the new factor $Z$
from the velocity-dependent term.
There is a weak $k$ dependence due to the fact that different
scales cross the horizon at different times, and hence
at different values of $\phi$.  
In particular, scale with the smallest $k$ crosses
the horizon first, and hence the corresponding $\phi$ is larger.
Whether the spectrum is tilted to the blue or to the red depends
on the form of $Z$.

By definition, the spectral index
\begin{equation}
n-1 = \frac{d \log |\delta_k|^2}{d \log k}
\end{equation}
where $n$ is in general a function of $k$.
We can equivalently 
express the spectral index $n$ in terms of the number of e-foldings $N$
when the scale $k$ crosses the horizon. One can show that 
\begin{equation}
n-1 = 2 \eta - 6 \epsilon + 2 \frac{d \ln Z}{d \ln k}
\end{equation}
Since $d \ln k = d \ln (aH)  \sim d \ln a = H dt$,
\begin{equation}
\frac{d \ln Z}{d \ln k} = \frac{1}{Z} \frac{d Z}{d \ln k}
= \frac{1}Z \frac{d Z}{d \phi} \frac{d \phi}{d \ln k}
= \frac{1}Z \frac{d Z}{d \phi} \frac{\dot{\phi}}{H}
=  \frac{1}Z \frac{d Z}{d \phi} 
\left( - \frac{M_P}{Z} \sqrt{2 \epsilon} \right)
= - \sqrt{2 \epsilon} \left( M_P \frac{Z^{\prime}}{Z^2} \right) 
\end{equation}
Let us define the parameters $\lambda$ and $\kappa$:
\begin{equation}
\lambda = M_P \frac{Z^{\prime}}{Z}, \quad \quad  
\kappa = M_P^2 \frac{Z^{\prime\prime}}{Z}
\end{equation}
Unlike the slow-roll parameters, $\lambda$, $\kappa$ are not required to 
be smaller than $1$. We have assumed
that $|\frac{1}{2} Z^{\prime} \dot{\phi}^2| << V$ in deriving
the equation of motion, this implies 
\begin{equation}
|\sqrt{2 \epsilon} \lambda| << Z
\end{equation}
Since $Z$ is of order $1$, and $\epsilon << 1$, it is easy to satisfy
the above condition.
To summarize:
\begin{equation}\label{index}
n-1 = 2 \eta - 6 \epsilon - \frac{2}{Z} \sqrt{2 \epsilon} \lambda
\end{equation}
Therefore, whether the velocity-dependent term tilts the spectrum
to the red or to the blue depends on the sign of $\lambda$.
If $Z$ has the same functional depenence on $\phi$ as $V$, then
$\lambda \sim \sqrt{\epsilon}$.
Hence, the velocity-dependent term has the same $k$ dependence as
$\epsilon$. This is the case if $V$ is due to higher dimensional gravity, 
where the post-Newtonian approximation \cite{weinberg} implies that 
\begin{equation}\label{postNewton}
Z(\phi) \approx 1 - V(\phi)/M_s^4, 
\end{equation}
where $M_s$ is the higher dimensional Planck mass or the string scale.
Since $V>0$ to provide inflation, $Z<1$, an important property of brane
inflation. If $Z$ is too small (not close to unity), higher terms in the 
effective action must be included.
In string theory,
when branes are moving with respect to each other, there is
a velocity-dependent potential of the form
${\cal V} = C \dot{\phi}^p/{\phi^q} + constant $,
where $C$ is a model-dependent constant, 
$p$ and $q$ are positive integers. 
We expect that $p$ is even by time reversal invariance.
In general, if supersymmetry is broken, 
the $p=2$ term is non-vanishing.
If, on the other hand, $Z(\phi)$ and $V(\phi)$ takes
different functional form, then the contribution of
the velocity-dependent potential to the spectral index can be bigger
or smaller than $\epsilon$. 

Another quantity of interest is the $k$ dependence of $n$. It is
also modified to:
\begin{equation}\label{ktilt}
\frac{d n}{d \ln k} = \frac{1}{Z} \left( -2 \xi + 16 \epsilon \eta -24 \epsilon^2 \right) 
+\frac{1}{Z^2} \left( 4 \epsilon \kappa - 8 \epsilon \lambda^2
+ 2 \eta \sqrt{2 \epsilon} \lambda   - 4 \epsilon \sqrt{2 \epsilon} \lambda \right)
\end{equation}
where
\begin{equation}
\xi = M_P^4 \frac{V^{\prime}V^{\prime\prime\prime}}{V^2}
\end{equation}
Typically, $\xi$ is the dominant contribution. 
Here, we note that even though the effects of $Z$ is small in
the spectral index $n$, it can have a measurable effect 
on ${dn}/{d \ln k}$.

To express the above in terms of $N$, we first
need an expression of $\phi_N$ as a function of $N$.
It takes $N$ e-foldings
for
$\phi_N$ to reach $\phi_{end}$. 
\begin{equation}
 N= \int_{t_N}^{t_{end}} H dt=
\int_{\phi_{N}}^{\phi_{end}} H \frac{d \phi}{\dot{\phi}} 
=-\int_{\phi_N}^{\phi_{end}} \frac{ 3H^2 Z d \phi}{V^{\prime}}
=-\frac{1}{M_P^2} \int_{\phi_N}^{\phi_{end}} \frac{ V }{V^{\prime}}
Z d \phi
\end{equation}
The value of $\phi_{end}$ is determined when the slow-roll condition 
breaks down. Typically, $\epsilon<\eta$,
as we will see in the examples,
and so the slow-roll condition breaks down when $\eta \sim 1$.

\section{Brane Inflation}

In the brane inflationary scenario \cite{dvali-tye}, the inflaton potential
is generated by the interaction between branes. This is naturally
realized in Type I string theory since its classical vacua
contain D-branes and orientifold planes \cite{Polchinski}. 
When the theory is compactified, the RR charges carried by the D-branes
are canceled by that of the orientifold planes, which must sit at 
orbifold fixed points. 
An ordinary orientifold plane ${\cal O}$
(corresponding to the worldsheet
parity transformation $\Omega$) carries negative RR charges
as well as negative tension (NS-NS charges).
When the D-branes are on top of an orientifold plane, the total energy
density and RR charges cancel. On the other hand,
an anti-orientifold plane $\widetilde{\cal O}$ carries positive 
RR charge and negative tension,
and so it cancels the energy density as well as RR charges of
anti D-branes. An ${\cal O}_{+}$ plane corresponding to a
frozen $D_8$ singularity \cite{Witten} in F theory
(and hence does not generate gauge symmetry)
carries positive RR charge
and positive tension.
More generally, there are four different types of orientifold planes 
(which carry positive/negative tension, and positive/negative RR charges 
respectively) corresponding
to four different orientifold 
projections \cite{ST,Sagnotti,Witten}.
If the internal manifold is smooth so there are no orientifold
planes where the curvature is localized, F theory provides
a proper description of such vacua.

We consider brane inflation when the extra dimensions are compactified,
so that the four-dimensional Planck scale is finite.
In the early universe, the branes and the orientifold planes do not have
to be exactly on top of one another. 
For parallel and static $D$-branes, and before supersymmetry is broken,
there is no force between the branes. 
However, if supersymmetry is broken, we expect a
static potential between the branes.
This is the case, for example, when
branes are not exactly parallel.
In general, the interbrane separations are parametrized by a matrix-valued
scalar field ${\bf \Phi}$. The diagonal values of ${\bf \Phi}$
correspond to the location of the branes.
The problem is usually quite involved if branes are intersecting at 
angles ({\it e.g.}, the brane configuration in \cite{ST}), 
as briefly discussed in \cite{Halyo}.
It is convenient to simplify the problem 
by considering a single brane or a stack of branes moving together.
A simplistic potential has the form
\begin{equation}
V (r) = \left( T_1 + T_2 \right) \pm M_s^{6-d_{\perp}} r^{2-d_{\perp}} \left( 1 + \sum_{i} e^{-m_i r} 
- \sum_{j} e^{-m_j^{\prime} r} \right) 
\end{equation}
where $d_{\perp}$ is the number of dimensions transverse to the branes,
$m_i$ and $m_j^{\prime}$ are the masses of the NS-NS and the RR states
respectively. Here,
$T_{1,2}$ are the tension of the two stacks of branes (and/or orientifold
planes)
respectively. For simplicity, we consider $3$-branes, although the analysis 
can be trivially generalized to arbitrary $p$-branes.
The $-$ sign corresponds to $T_1 T_2 >0$.
The $+$ sign corresponds to branes with $T_1 T_2 <0$ since
the potential between the branes is repulsive if
they have opposite tensions. 
Note that there is no (light) open strings ending on orientifold planes. So
there is no brane mode associated with brane-orientifold plane separations.
So here, one should view a negative $T_1$ as due to a brane sitting 
on top of an orientifold plane.
For $d_{\perp}=2$, the potential is not inverse-power like but logarithmic. 
In any case, for sufficiently large $r$, the potential
$V(r)$ is flat.
Very crudely, the magnitude of the density perturbation is 
$\delta \rho/\rho = |\delta_k| \approx M_s/M_P$. 
In the brane world scenario,
both $M_s$ and the effective Planck scale $M_P$ during inflation
can take a wide range of values.
To obtain the correct 
magnitude, typically $M_s \approx 10^{12}$ GeV. If $M_s$ is much smaller, then 
the size of compactification must vary such that the effective Planck 
scale during inflation is substantially smaller than that at present. 
The dynamics of 
such a radion mode is discussed in Ref\cite{ira}. Here we shall simply take
$M_s \approx 10^{12}$ GeV.

One may also consider (non-supersymmetric) models with both branes and 
anti-branes \cite{Dvali,Burgess,Halyo}.
The potential for small brane-anti-brane separation 
${\bf{r}} - {\bf{r}}_i$ has the form
\begin{equation}
\label{rmcase}
V( {\bf{r}}) = A - \frac{B_i}{| {\bf{r}} - {\bf{r}}_i|^m}
\end{equation}
where $m=d_{\perp}-2$. 
This is the case considered in Ref\cite{Dvali,Burgess}.
For generic values of $A$ and $B_i$ given by string theory, 
the slow-roll conditions cannot be satisfied due to the physical 
requirement that $r = |{\bf{r}} - {\bf{r}}_i| < r_{\perp}$, 
where $r_{\perp}$ is the extra 
dimension compactification size, and the relative large value of 
$B$\cite{Burgess}. (We note here that for $d_{\perp}=2$,
{\it a priori}, it is possible to have 
$\eta <1$. Unfortunately, the allowed values of $\eta$ will give rise to 
a tilt too large to be in agreement with the
observational results $ |n-1| <  0.1$.)
However, depending on the distribution of branes and orientifold planes, 
there are regions in the extra dimensions in which the potential may be
sufficiently flat. For example, if the branes are close to an 
orientifold plane, the tension is screened; to the anti-brane, the 
brane-orientifold plane system 
behaves like a dipole, so $B_i$ is essentially the dipole moment and 
$m=d_{\perp}-1$ (or more generally, higher moment if the dipole moment
vanishes). 
In this case, $B_i$ can be made as small as one wants
to satify the slow-roll conditions, and the analysis in Ref\cite{Dvali} 
is applicable. In reality, the separation
between the brane and the orientifold plane (that is, the dipole moment $B_i$)
should also be treated as a dynamical field.

When the brane separation $r$ is comparable to the the compactifaction
radius, the contribution of the image charges (dipoles)
become important:
\begin{equation}
V( {\bf{r}}) = A - \sum_{i} \frac{B_i}{| {\bf{r}} - {\bf{r}}_i|^m}
\end{equation}
where the sum is over all the branes and orientifold planes in the covering
space. In general, the value of $m$ can even be different
for different clusters of branes and orientifold planes.
A few comments on this potential
are in order. First of all, the sum may be divergent, for example,
when all $B_i$ have the same sign as in \cite{Burgess}.
One should regulate the above sum so that it is finite.
After the regularization, the sum becomes finite but can still be non-zero
if the sum of tensions of the branes do not cancel. 
The non-zero constant can renormalize
the value of $A$, thereby changing the predictions of the power spectrum.

Following Ref\cite{Burgess}, let us suppose that at ${\bf r} ={\bf r}_0$, 
the force from various branes and orientifold planes cancel.
We expand $V({\bf r})$ for small $z \equiv |{\bf r} - {\bf r}_0|$,
The linear term vanishes since the force is zero at ${\bf r}_0$.
In general, the $z^2$ term is non-zero.
However, in some special cases, the compactification implies
certain discrete symmetries which forbid the quadratic
terms to contribute. For example,
in the case of ${\bf Z}_4$ 
orientifold, the orbifold symmetry requires the compactified
dimensions be a product of two-dimensional square lattices.
For concreteness, let us take ${\bf r}_0$ to be the center of the lattice,
and the potential has the form
\begin{equation}
\label{z4case}
V(r) = A - C z^4  + \dots
\end{equation}

During inflation, $z$ may start out small, described by $V(z)$ (\ref{z4case}), 
but end up with small ${\bf{r}} - {\bf{r}}_i$, described by V(r) 
(\ref{rmcase}). This means that the correct potential $V(r)$ should have 
the form:
\begin{equation}\label{interpolateV}
V(r) = \nu M_s^4 \left[ 1  - \frac{\zeta}{4}
 \frac{z^4}{r^m} f(r)  
\left( 1 + \sum_i e^{-m_i r} - \sum_j e^{-m_j r} \right) \right]
\end{equation}
where $\nu$ is a dimensionless number of order $1$, and $f(r)$ is
a smooth function. The precise form of $f(r)$ depends on the
distribution of the branes.
In reality, since the compactification (the lattice)
breaks spherical symmetry,
V({\bf r}) is a function of the vector ${\bf r}$ (and not simply
$r$).
This is similar to
hybrid inflation since there are $d_{\perp}$ number of fields. 
The set of inflatons can wander around before falling onto one of the branes. 
In other words, 
different paths that the inflaton fields follow
will yield different numbers of e-foldings. This resembles hybrid inflation.

This potential interpolates from $V(z)$ at small $z$
to $V(r)$ when the anti-brane is close to the brane. Finally, when $r$ is even
smaller, the stringy modes begin to contribute. Therefore,
inflation ends sharply since the exponentials
are no longer small. When a stack of $N_p$ $p$-branes
and a stack of $N_a$ anti-$p$-branes annihilate,
we may end up with (for $N_p>N_a$) a stack of $(N_p-N_a)$ $p$-branes, or
if $N_p=N_a$, we may end up with $(p-2)$-branes (depending on the
details of the situation).

As the form of the potential changes as the branes move
(since they are at different locations),
different physical scales sample a different form of the inflation potential
since 
they cross the horizon at different time.
The density perturbation spectrum can be tilted
rather differently for small and large scales. The precise form
depends on 
the distribution of the
branes, as well as the path that the branes follow.
 
As the branes are in relative motion, generically supersymmetry
is broken. Therefore, in addition to the static potential,
there is a velocity-dependent potential \cite{Bachas}
(see also \cite{Polchinski} for a detail discussion).
Generically, the velocity-dependent potential has the same form as the static potential.
Assuming that the density perturbation were generated before 
the stringy modes become important, $Z(\phi)$ takes
the form
\begin{equation}
Z = 1 - \frac{\widetilde{\zeta}}{4} \frac{z^{4}}{r^m} f(r)
\end{equation}
Following the general analysis in the previous section,
we see that its contribution to the tilt is of the same order as $\epsilon$.

\section{Examples}

\subsection{An Interpolating Potential}

Let us study the effects on density perturbation from (\ref{interpolateV}).
In general, $m \leq d_{\perp}-2$ since there is cancellation between
positive and negative tension objects. In particular, $m=d_{\perp}-1$
for dipoles. 
For simplicity, we consider $m=d_{\perp}-2$. 

We can see from Eq.(\ref{T/S}) that 
the amplitude of the tensor to scalar perturbations
is modified by a factor of $1/Z^2$, whose value 
is model-dependent. 

Let us consider a potential of the form
\begin{equation}\label{full}
V(r) = \nu M_s^4 \left( 1 -  \frac{\zeta}{4} \frac{z^4}{r^{d_{\perp}-2}} 
\right)
\end{equation}
For small $r$, $z \sim \sqrt{d_{\perp}} r_{\perp}/2$, and so
$V(r) \sim A - {B}/{r^{d_{\perp}-2}}$. To compare with \cite{Burgess}:
\begin{equation}
\zeta = 2 \left(\frac{2}{\sqrt{d_{\perp}}} \right)^4 \alpha \beta g_s
M_s^{2-d_{\perp}} r_{\perp}^{-4}
\end{equation}
where $g_s$ is the string coupling, and $\beta$ is given by
\begin{equation}
\beta = \pi^{-d_{\perp}} \Gamma(\frac{d_{\perp}}{2}-1)
\end{equation}
On the other hand, for small $z$, 
$r \sim r_{\perp}/2$ and hence 
\begin{equation}\label{smallz}
V(r) \sim A - \frac{C}{4} z^4
\end{equation}
where
\begin{equation}
C= \left(\frac{2}{\sqrt{d_{\perp}} r_{\perp}} \right)^{d_{\perp}-2} \zeta
= 2 \left(\frac{2}{\sqrt{d_{\perp}} r_{\perp}} \right)^{d_{\perp}+2}
\alpha \beta g_s
M_s^{2-d_{\perp}}
\end{equation}

\begin{figure}[tb]
\begin{center}
\begin{tabular}{c}
\epsfxsize=5cm
\epsfbox{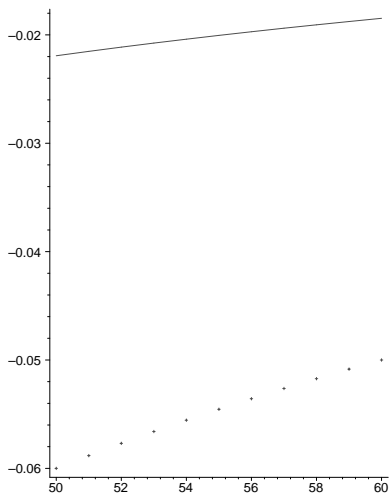} 
\end{tabular}
\end{center}
\caption[]{The tilt of the density perturbation spectral index $n-1$ 
for $d_{\perp}=6$ as 
a function of the number of e-foldings $N$ from the end of inflation.
The solid line corresponds to the full potential of the form
$1- \zeta z^4/{4r^{d_{\perp}-2}}$.
The set of points correspond to $-3/N$, the tilt for the approximate
potential $ A- C z^4/4$. \label{tilt}}
\end{figure}

Let us define 
\begin{equation}
\gamma = 4 \beta \left(\frac{2}{\sqrt{d_{\perp}}}\right)^{d_{\perp}+2}
\end{equation}
Suppose we can approximate the potential by (\ref{smallz}) all the way to
the end of slow-roll, then
\begin{equation}
\eta = - 3 \gamma \left(\frac{z}{r_{\perp}}\right)^2
\end{equation}
The slow-roll ends when 
$z_{end} = r_{\perp}/{\sqrt{3 \gamma}}$. 
For example, if $d_{\perp}=6$, $z \sim 3.6 r_{\perp} \sim 3
\sqrt{d_{\perp}} r_{\perp}/2$. Therefore, before the end of slow-roll
is reached (and before tachyonic
instability is developed since it was assumed 
that $M_S r_{\perp} >> 1$ in \cite{Burgess}), 
the approximation (\ref{smallz})
breaks down.
Moreover, since the branes almost collide at the end of inflation,
there is not enough time for the branes to reheat.
One should use the full inflaton
potential (\ref{full}).

It is useful to define
$y=2z/(\sqrt{d} r_{\perp})$, so that $y_{end} << 1$ implies that
the approximation (\ref{smallz}) is valid.
The slow-roll parameter $\eta$ is
given by
\begin{equation}
\eta \sim - 4 \beta \left(\frac{2}{\sqrt{d_{\perp}}}\right)^{d_{\perp}}
\frac{y^2}{\left(1-y\right)^{d_{\perp}}}
\left[ \left( 3 + \left(d_{\perp}-4\right) y \right) 
\left(1+\left(\frac{d_{\perp}}{4}-\frac{3}{2}\right)y \right) 
+ y\left(1-y\right)\left(\frac{d_{\perp}}{4}-\frac{3}{2}\right) 
\right]
\end{equation}
The slow-roll parameter $\epsilon \sim y^6$ and so
$\epsilon << \eta$. Hence, we will ignore 
it in what follows. As discussed, the velocity-dependent term
has the same form as the static potential, and so its contribution to
the spectral index is also negligible.

The number of e-foldings is
\begin{equation}
N= \frac{1}{4 \beta  \left(\frac{2}{\sqrt{d_{\perp}}} \right)^{d_{\perp}}} ~
\int_{y_{N}}^{y_{end}} \frac{(1-y)^{d_{\perp}-1} dy}{y^3 \left( 1 + (\frac{d_{\perp}}{4}-\frac{3}{2}) y \right)}
\end{equation}

The tilt in the spectral index as a function of $N$ for $d_{\perp}=6$
is depicted in Figure \ref{tilt}.
With the full potential (\ref{full}),
the value of $z_{end}$ at the end of slow-roll is smaller than the
naive answer $r_{\perp}/\sqrt{3 \gamma}$. 
Therefore, there is enough time to reheat, 
without having to invoke tachyonic instability.
However, the initial condition of $z_N$ at $N=60$
is also smaller with the full potential. Hence the branes has to be closer
to $z \sim 0$ in the beginning of inflation.

\subsection{Inverse Power Potential}

A potential of the form
\begin{equation}
V(\phi) = \nu M_s^4 \left( 1 - \frac{c}{\phi^m} \right)
\end{equation}
was considered in \cite{Dvali,Burgess}
when branes and anti-branes collide.
The slow-roll parameters, 
\begin{equation}
\epsilon = \frac{1}{2} M_P^2 \frac{m^2 c^2}{\phi^{2m+2}}, \quad \quad
\eta = - M_P^2 \frac{m(m+1)c}{\phi^{m+2}}
\end{equation}
As discussed, $m$ does not have to be $d_{\perp}-2$; for
dipoles, $m=d_{\perp}-1$.
Here, we can phenomenologically model the behavior of the potential
with an arbitrary $c$ since its precise value depends on the brane
distribution.
The slow-roll condition breaks down when $\eta \sim 1$:
\begin{equation}
\phi_{end} \sim \left( M_P^2 c m(m+1) \right)^{1/(m+2)}
\end{equation}

Let us consider a velocity-dependent potential:
\begin{equation}
Z = 1 - \frac{\tilde{c}}{\phi^m}
\end{equation}
then
\begin{eqnarray}
N &=&  -\frac{1}{M_P^2} \frac{1}{m c} \int_{\phi_N}^{\phi_{end}}
\phi^{m+1} \left( 1 - \frac{\tilde{c}}{\phi^m} \right) 
d \phi \nonumber \\
&=&
\left(\frac{m+1}{m+2} \right)
\left[ \left( \frac{\phi_N}{\phi_{end}}
\right)^{m+2} - 1\right]  
- \frac{\tilde{c}}{\phi_{end}^m} \left(\frac{m+1}
{2}\right) 
\left[ \left(\frac{\phi_{N}}{\phi_{end}}\right)^{2} - 
1 \right]
\end{eqnarray}
The solution of $\phi_N/\phi_{end}$ does not change greatly with
the velocity-dependent term,
\begin{equation}
\left(\frac{\phi_N}{\phi_{end}}\right)^{m+2} \sim 
\left(\frac{m+2}{m+1} \right) N
\end{equation}
Furthermore, $\epsilon << \eta$ and so the velocity-dependent term
is also negligible.
Therefore,
\begin{equation}
n-1 \sim - \frac{2}{N} \left( \frac{m+1}{m+2} \right)
\end{equation}
where $m \leq d_{\perp}-2$. The spectrum is tilted to the red.
If the lowest non-vanishing contribution
to the potential comes from the dipole of 
a brane-orientifold plane system, $m= d_{\perp}-1$ which gives a
slightly larger red tilt.
 
\section{Discussion}

In this paper, we have examined the density perturbation spectrum in
the brane inflationary scenario. We have also studied
the effect of a velocity-dependent
potential. The form of the inflation potential
as well as the velocity-dependent potential
is strongly motivated from higher dimensional 
gravity (in particular, string
theory). In the usual four-dimensional inflationary scenario,
the deviation of $Z(\phi)$ from unity is due to quantum effects
and is usually small. In brane inflation, however,
the velocity-dependent potential can give rise to a significant
effect in $Z(\phi)$.  
Generically, the effect on the spectral index is comparable
with that from the slow-roll parameter $\epsilon$ (from Eq.(\ref{index})).
Therefore, in models where $\epsilon$ is not small, this may be
a measurable effect. Even though this effect on $n$ can be small
in a generic model, the scale-dependence of $n$ has a stronger
dependence on the velocity dependent term, as it scales 
$dn/d \ln k$ by a factor of $1/Z$ (see Eq. (\ref{ktilt})).
More importantly, the velocity-dependent
potential can also modify the ratio of tensor to scalar fluctuations
(see Eq.(\ref{T/S})). 
With a global analysis of data
from various measurements, one should be able to distinguish
these effects from those that can be obtained from
a conventional inflationary scenario.
We expect that a few percent deviation of $Z$ from unity may be
measurable.

It seems that the power spectrum index in a generic inflationary scenario
in the brane world has a red tilt of a few percent.
The velocity dependent potential can further tilt the density perturbation
spectrum to the red or to the blue, depending on the specific form
of the potential. The blue tilt due to the velocity-dependent term 
does not require extra fields, in contrast to other inflationary 
models ({\it e.g.}, hybrid
inflation).

The velocity-dependent potential in the brane flationary scenario
is similar to that appears in k-inflation \cite{k-inflation}. However, unlike
k-inflation, in our scenario, inflation is still driven by the static
potential. The additional velocity dependent term provides
modifications of the usual slow-roll inflation. Furthermore, the kinetic
term and the static potential in the brane inflation scenario are related
by string theory or by post-Newtonian approximation 
(see Eq.(\ref{postNewton})).

Potential of the type $1/\phi^m$ has been studied in the
context of quintessence \cite{inverse}. 
It was argued in \cite{binetruy}
that this type of potential appear naturally in
supersymmetric theory. As we have described,
this type of potential is also motivated when
we study cosmology involving branes.
Furthermore, the velocity dependent term has a close resemblance
to that in k-essence \cite{k-essence}. It would be interesting to explore
whether the specific form of k-essence motivated by brane dynamics may
offer an explanation for the present accelerating universe.

\acknowledgments

The research of G.S. was supported in part by
the DOE grant FG02-95ER40893, the University of Pennsylvania
School of Arts and Sciences
Dean's fund, and the NSF grant PHY99-07949. 
The research of S.-H.H.T. was partially supported by the National
Science Foundation.
G.S. would like to thank
the M-theory workshop at ITP, Santa Barbara, and 
the Theory Division at CERN
for support and hospitality during the
final stage of this work.

\appendix

\section*{Quantum Fluctuations}\label{appendix:quantum}

In this appendix, we derive the equation for the quantum
fluctuation $\delta \phi$ in the presence of a non-trivial $Z(\phi)$, 
and show that at late time,
$\delta \phi$ and $\dot{\phi}$ obey the same 
differential equation with respect to time.
We used this fact to derive Eq.(\ref{gpformula})
analogous to the original derivation of \cite{GuthPi}.

To study quantum fluctuations, we split $\phi$ into the classical piece
$\phi_c$ (which satisfies (\ref{eom})) and the quantum piece $\delta \phi$:
\begin{equation}
\phi = \phi_c + \delta \phi
\end{equation}
The classical piece is homogeneous, whereas the fluctuations satisfy:
\begin{equation}\label{quantum}
Z \left( \delta \ddot{\phi} + 3 H \delta \dot{\phi} \right) 
+ Z^{\prime}  \left( \ddot{\phi} + 3 H \dot{\phi} \right)  \delta \phi
- \frac{Z}{a^2} \nabla^2 \delta \phi
+ V^{\prime \prime} \delta \phi = 0
\end{equation}
where we have dropped the subscript $c$ for the classical part of $\phi$.
We note that for scales outside the horizon (hence the gradient term
is negligible), $\delta \phi$ and $\dot{\phi}$ obeys the same
time-dependent differential equation.
By expanding $\delta \phi$ in fourier modes:
\begin{equation}
\delta \phi (x,t) = \int \frac{d^3 k}{\sqrt{2 k}} \left( a_{\bf k} 
\delta \phi_{\bf k} (t) 
e^{i {\bf k} \cdot {\bf x}} 
+ a_{\bf k}^{\dagger} \delta \phi_{\bf k}^{*} (t) e^{-i {\bf k} \cdot {\bf x}} \right)
\end{equation}
The equation of motion for $\delta \phi_{\bf k}$ is
\begin{equation}
Z \left( \delta \ddot{\phi}_k + 3 H \delta \dot{\phi}_k \right) 
+ Z^{\prime}  \left( \ddot{\phi} + 3 H \dot{\phi} \right)  \delta \phi_k
+ \left( Z \frac{k^2}{a^2} + V^{\prime \prime} \right) \delta \phi_k= 0
\end{equation}
It is convenient to rewrite it in terms of $u_{\bf k}=a \delta \phi_{\bf k}$.
In the conventional case, $Z (\phi)=1$, 
\begin{equation}
u_{\bf k}^{\prime \prime} + \left( k^2 -\frac{a^{\prime \prime}}{a} 
+a^2 V^{\prime\prime} \right)
u_{\bf k} = 0
\end{equation}
where prime denotes derivative with respect to conformal 
time $\eta=-H^{-1}e^{-Ht}$. For de Sitter space, the solution is
\begin{equation}\label{u_k0}
u_{\bf k} (\eta) = \frac{1}{2} \sqrt{-\pi \eta} H_{3/2} (- k \eta)
\end{equation}
which reduces to $u_{\bf k} \rightarrow \frac{1}{\sqrt{2k}} e^{- i k \eta}$
for $k>> a H$ and $u_{\bf k} \sim a$ for $k<<a H$.

In the presence of a velocity-dependent potential,
\begin{equation}
u_{\bf k}^{\prime \prime} + \left( k^2 -\frac{a^{\prime \prime}}{a} 
+ \frac{a^2}{Z} V^{\prime\prime} -\frac{a^2 V^{\prime} Z^{\prime}}{Z^2} \right)
u_{\bf k} = 0
\end{equation}
For scales inside the horizon, when the fluctuations are 
generated, the dominant term is still $k^2$. Therefore, $u_{\bf k}$
is still given by the Bunch-Davies vacuum,
\begin{equation}
u_{\bf k} = \frac{1}{\sqrt{2k}} e^{-i k \eta}
\end{equation}
Hence the quantum fluctuations $\delta \phi_{\bf k}$ is still
given by the de Sitter temperature,
\begin{equation}
\delta \phi_{\bf k} = \frac{H}{2 \pi}
\end{equation}

\end{document}